\documentclass[a4paper]{article} 
\pdfoutput=1

\usepackage{times}
\usepackage{natbib}
\usepackage{graphicx}
\usepackage[table]{xcolor}
\usepackage{makecell}
\usepackage{soul}
\usepackage{xcolor}
\usepackage{multirow}
\usepackage{amsmath}
\usepackage{amssymb}

\title{Reviewing the role of the extinction coefficient \\in radar remote sensing}
\author{J. David Ballester-Berman
\\ \vspace{0.3cm}\small{Universitat d'Alacant, Spain}\\
\small{\today} }
\date{}

\definecolor{red_light}{rgb}{0.88,0.5,0.5}
\definecolor{grey}{rgb}{0.55,0.55,0.55}

\begin{document}

\maketitle

\pagestyle{headings}
\setcounter{page}{1}
\pagenumbering{arabic}

\begin{abstract}
This report revisits the role of the extinction coefficient in radar backscattering-based models for forest monitoring. A review of a number of works dealing with this issue has revealed a diversity of extinction values being unclear its dependence on the sensor frequency and the forest type. In addition, a backscattering model directly derived from the RVoG formulation is employed to analyse the saturation of backscattering level as a function of vegetation height and the presence of a decreasing trend of backscatter beyond the saturation point as suggested in previous works in the literature. According to this analysis it seems reasonable to think that further research specially focused on dedicated experimental measurements of the extinction coefficient should be carried out.
\end{abstract}

\section{Introduction and Motivation}

The retrieval of forest height and biomass by radar remote sensing has become a mature technique in recent years. The early empirical approaches are based on regression analyses relating backscattering power to forest height/biomass previously known by other means. However, this strategy presented the well-known saturation effect \cite{ar:imhoff1,ar:imhoff2}. The transition to a more advanced framework based on multidimensional SAR approaches has been possible by combining polarimetric and interferometric information. Hence, interferometric- (from early works by Treuhaft et al. \cite{ar:treuhaft,ar:treuhaft_siqueira} to very recent contributions such as \cite{ar:lei2019,b:handbook,ar:shiroma2020,ar:lei2021}) and tomographic-based radar techniques (from the first airborne demonstration \cite{ar:reigber00} to integral approaches for forest characterization \cite{ar:tello2018,ar:Aghababaee2018,ar:dalessandro2019}) have become the current tools for forest structure understanding and monitoring.

Despite these technical and methodological advances, the saturation of radar backscattering power from forests still remains to be fully understood as discussed in many papers along the past and recent years employing SAR data acquired at P-, L-, and C-band. Some works focused on identifying the saturation point of radar backscatter \cite{ar:hensley2014} by exploiting a functional form relating backscatter and biomass in order to find the range of biomass values that can be estimated in a reliable way. Yu and Saatchi \cite{ar:yu_saatchi} extensively and globally investigated the L-band backscatter relationship to biomass across eleven different forest types showing that each particular biome is characterised by different backscatter-biomass relationship and, hence, a different associated saturation level. Authors employed empirically-based radar backscatter models and showed that L-band radar saturation level is greater than 100 $\mbox{Mg}/\mbox{ha}$ on the average with boreal forests and temperate conifers having an enhanced sensitivity to forest biomass reaching values higher than 200 $\mbox{Mg}/\mbox{ha}$. This fact had already been demonstrated by Santoro et al. in \cite{ar:santoro2011} for boreal forest growing stock volume by employing multi-temporal C-band data sets. The combination of multi-temporal and multi-frequency approaches has been also successfully tested as shown by Englhart et al. \cite{ar:englhart2011}. In this line, more recently, Cartus et al. \cite{ar:cartus2019} have also reported improvements on tropical forest biomass retrieval by combining multi-temporal and multi-frequency acquisitions and Hayashi et al. \cite{ar:hayashi2019} investigated the enhancement of the saturation point in dense tropical forest up to 280 $\mbox{Mg}/\mbox{ha}$ by using L-band ALOS-2 times series SAR data. An early and detailed work on the saturation point effect for temperate forests was presented by Watanabe et al. in \cite{ar:watanabe2006}.

However, investigations by Mermoz et al. \cite{ar:mermoz2015} on the relationship between L-band backscatter and dense tropical forest biomass have found that a negative linear correlation appears after reaching the maximum backscatter. Authors in \cite{ar:mermoz2015} analysed HV ALOS-PALSAR images and experimentally observed this increasing-decreasing behaviour of backscatter which was also supported by a simulation study based on a detailed electromagnetic backscatter theoretical modelling. The reason for such an effect was found in the total wave extinction which increases as the forest biomass does. The decreasing linear trend beyond saturation point for dense tropical forests seems to be either confirmed or not depending on the site under study. Although not explicitly stated by authors in \cite{ar:yu_saatchi}, the reported ALOS HV backscattering signatures corresponding to tropical rainforest in Latin America, Africa and Asia exhibit a decreasing trend beyond the saturation point (see Fig. S3 in \cite{ar:yu_saatchi}) in agreement with the findings by Mermoz et al. \cite{ar:mermoz2015}. On the other hand, this behaviour has not been reported by other works such as in \cite{ar:cartus2019} or \cite{ar:rodriguez2019}. In \cite{ar:cartus2019} authors perform a backscatter sensitivity analysis for multiple frequencies in tropical forests and no evidences on the trend change beyond the saturation point are seen.


In line with the suggestions made in \cite{ar:mermoz2015}, a work by Joshi et al. \cite{ar:joshi2017} unveiled some observations regarding the impact of different forest structural parameters, i.e. stem height and size, density or presence of understory on the backscattered radar signal. That study was focused on conifer and broadleaf forests in Spain and Denmark with moderate heights up to about 20 m. Based on a number of evidences, it suggests that a prior knowledge of vegetation cover fraction or stem number densities would allow the use of backscattered-derived above ground volume estimates beyond the saturation point, even though further research is also recommended.

Since the original Water Cloud Model by Attema (1978) and the extended version accounting for non-spherical particles by Ulaby (1986), the construction of both physically-based and empirical backscatter models relies on the extinction coefficient as it controls the variation rate of backscattering as a function of height and biomass. However, a review of the past and recent literature reveals that the role of the extinction coefficient in the backscatter response is not fully understood yet as explicitly stated in \cite{ar:mermoz2015} and more recently in \cite{ar:cartus2019}. This is indeed a parameter with a clear physical meaning as it has a direct relationship with the physical scattering process involved in radar signals interaction with vegetation. However, as stated above it is sometimes considered as a kind of fitting parameter rather than a physically measurable indicator, and it actually could take different values depending on the selected area of study and sensor configuration as shown in a number of works. 

The objective of this report is twofold. First, a survey on the works employing, estimating and/or measuring 
the extinction coefficient is carried out in order to summarize how this parameter has been considered in the 
literature. It is noted that this survey has taken into account those contributions which devoted some effort 
to specifically assess the effect of extinction in physical parameter retrieval for forest scenarios. Second, we revisit its role on the theoretical backscattering response by making use of a backscattering physically-based model whose 
formulation is directly derived from the Random Volume over Ground (RVoG) model by Treuhaft and 
co-workers \cite{ar:treuhaft,ar:treuhaft_siqueira}. There is no need to mention the contribution of this 
model to SAR interferometry and polarimetric SAR interferometry. Nevertheless, it is also possible to 
straightforwardly derive the corresponding version for backscattering analyses being a model whose 
formulation only relies on physical foundations through the distorted Born approximation. Essentially, we 
employ the same model as Hensley et al. did in \cite{ar:hensley2014} but keeping the physical meaning of all 
different parameters involved in original formulation proposed for the RVoG \cite{ar:treuhaft_siqueira}.

\section{The extinction coefficient in forest scenarios}

Next, a compilation of the works focused on forest parameter estimation providing some kind of elaborated observations either qualitative or quantitative or both on the role of the extinction coefficient is presented. This can be a helpful way to provide a general overview on the current knowledge on this parameter. The information is compiled in Tables \ref{t:ext1} and \ref{t:ext2}. Citations to the contributions are given together with the extinction values reported and the radar sensor frequency band. In addition, some comments retrieved from each particular analysis and highlighted by the corresponding authors are also pointed out. It is noted that grey coloured cells refer to works where the extinction values were experimentally measured by comparing backscattering levels from forests and trihedral corner reflectors. Some early pioneering works on backscattering and attenuation studies from vegetation such as \cite{ar:attenuation1} and others were not overlooked, despite not being included here as they do not provide extinction measurements or direct estimations from whole forest stands but from some particular components whose radar signature is studied in detail.

\begin{table}[h!]
\renewcommand{\arraystretch}{1.0}
\scriptsize
\begin{center}
\begin{tabular}{|l|c|c|c|c|c|}
\Xhline{1.5pt}  \rowcolor{red_light}
{\bf Reference} & {\bf Forest type} & {\bf Extinction value (dB/m)} & {\bf Polarisation} &
{\bf Freq. Band} & {\bf Comments} \\ 
\Xhline{1.5pt} Ulaby et al. (1990) \cite{ar:propro} & \shortstack{\\Red pine trees \\(dense canopy)} & 0.26 / 0.25  & HH / VV & L & \cellcolor{grey!45}\shortstack{\\ Backscattering \\measurements} \\
\hline 
\shortstack{\\ Pulliainen et al. (1994) \cite{ar:pulliainen} \\ \vspace{0.5cm}} & \shortstack{\\ Boreal\\ \vspace{0.5cm}} & \shortstack{\\ 0.08 \\ \vspace{0.5cm}}& \shortstack{\\ HH / VV \\ \vspace{0.5cm}}& \shortstack{\\ C and X\\ \vspace{0.5cm}} & \shortstack{\\ Semi-empirical model\\ Extinction can be derived from \\ regression coefficients reported \\ Marginal differences \\at both frequencies}\\
\hline 
\shortstack{\\ Fleischman et al. (1996) \cite{ar:fleischman96} \\ \vspace{1.3cm}}& \shortstack{\\ Boreal  \\ \vspace{1.3cm}} & \shortstack{\\ $\theta_0$=30$^\circ$, 0.07 / 0.09 \\ $\theta_0$=45$^\circ$, 0.07 / 0.13 \\ $\theta_0$=60$^\circ$, 0.06 / 0.1 \\ \quad \\ $\theta_0$=30$^\circ$, 0.19 / 0.18 \\ $\theta_0$=45$^\circ$, 0.18 / 0.19 \\ $\theta_0$=60$^\circ$, 0.17 / 0.17 \\ \quad \\ $\theta_0$=30$^\circ$, 0.42 / 0.4 \\ $\theta_0$=45$^\circ$, 0.43 / 0.39 \\ $\theta_0$=60$^\circ$, 0.33 / 0.32} & \shortstack{ \\ HH / VV \\ \vspace{1.3cm}} & \shortstack{ \\ \quad \\ UHF  \\ \quad \\ \quad \\ \quad\\ \quad \\ \quad \\ \quad \\ L \\ \quad \\ \quad \\ \quad\\ \quad \\ \quad \\ \quad \\C \\ \quad \\ \quad \\ \quad \\ \quad \\ \quad} & \cellcolor{grey!45}\shortstack{\\ Backscattering measurements \\ \\ Canopy height was measured \\ during field work but\\ 
it was not reported \\neither in \cite{ar:fleischman96} nor in \cite{ar:toups96} \\ We assume a 25 m forest height \\ \quad \\ \quad \\ \quad \\ \quad\\ \quad} \\
\hline 
\shortstack{\\Treuhaft et al. (1996) \cite{ar:treuhaft}\\ \vspace{0.3cm}} & \shortstack{\\Boreal\vspace{0.3cm}} & \shortstack{\\ frozen / thawed \\0.11  / 0.22  \\ 0.12  / 0.24  \\ 0.21  / 0.42 } & \shortstack{\\VV\vspace{0.3cm}} &\shortstack{\\ C\vspace{0.3cm}} & \shortstack{\\ InSAR studies\\ Derived from extinction values from \\ Kasischke et al. (1989) at C- and X-band \cite{ar:kasischke89} \\and backscattering analysis at L-band in \cite{ar:way1990}} \\
\hline 
\shortstack{\\Askne et al. (1997) \cite{ar:askne}\\ \vspace{0.25cm}} & \shortstack{\\Boreal\\ \vspace{0.25cm}} & \shortstack{\\0.53-0.57\\ \vspace{0.25cm}} & \shortstack{\\VV\\ \vspace{0.25cm}} & \shortstack{\\C\\ \vspace{0.25cm}} & \shortstack{\\ InSAR approach \\ Assumed value considering \\ previous work in \cite{ar:propro}, \cite{ar:pulliainen} and \cite{ar:fleischman96}} \\
%
%
\hline 
\shortstack{\\Treuhaft et al. (2000) \cite{ar:treuhaft_siqueira} \\ \vspace{0.3cm}}& \shortstack{\\Boreal \\ \vspace{0.3cm}}& \shortstack{\\0.3\\ \vspace{0.3cm}} & \shortstack{\\HH / VV\\ \vspace{0.3cm}} & \shortstack{\\C\\ \vspace{0.3cm}} & \shortstack{\\ Combined InSAR and backscattering \\ Assumed value \\ Estimation range is 0 to 0.4 dB/m \\but not considered for further analysis}\\
\hline 
\shortstack{\\Santoro et al. (2002) \cite{ar:santoro2002} \\ \vspace{0.25cm}}&\shortstack{\\ Boreal \\ \vspace{0.25cm}}& \shortstack{\\2 \\ \vspace{0.25cm}}& \shortstack{\\VV\\ \vspace{0.25cm}} & \shortstack{\\C\\ \vspace{0.25cm}} & \shortstack{\\ Combined InSAR and backscattering \\ Assumed value suggested in \cite{ar:dammert1999}, a research report \\ not found after online search} \\
\hline Cloude et al. (2003) \cite{ar:3stage} & Simulation  & 0.28 & Quad-pol & L & \shortstack{\\Assumed value after theoretical \\ analysis of RVoG model}\\ 

\hline \shortstack{\\ Askne et al. (2005) \cite{ar:askne2005} \\ \quad \\ \quad \\ \quad} & \shortstack{\\Boreal \\ \quad \\ \quad \\ \quad}& \shortstack{\\2 \\ \quad \\ \quad \\ \quad}& \shortstack{\\VV\\ \quad \\ \quad \\ \quad} & \shortstack{\\C\\ \quad \\ \quad \\ \quad} & \shortstack{\\ Combined multitemporal and IWCM \\ Assumed value based \\ on previous experience \cite{ar:santoro2002} \\ \emph{Not a critical value} } \\

\hline 
\shortstack{\\Thirion-Lefevre et al. (2006) \cite{ar:thirion_COSMO}\\ \vspace{0.05cm}} & \shortstack{\\Maritime  \\ pine trees} & \shortstack{\\0.1-0.2 / 0.15-0.25 \\ 0.2-0.3 / 0.2-0.25} & \shortstack{\\ HH / VV \\ HH / VV } & \shortstack{\\ P \\ L} & \cellcolor{grey!45}\shortstack{\\ Backscattering \\measurements}\\
\hline 
\shortstack{\\ Thirion-Lefevre et al. (2007) \cite{ar:thirion2007} \\ \vspace{0.1cm}}& \shortstack{\\ Simulation \\ Tropical and  \\ temperate} & \shortstack{\\0.15 / 0.23 \\0.2 / 0.23 \\ \vspace{0.1cm}} & \shortstack{\\ HH / VV \\ HH / VV \\ \vspace{0.1cm}} & \shortstack{\\ P \\ L \\ \vspace{0.1cm}} & \shortstack{\\ Std. dev. 0.04 / 0.05 dB/m \\Std. dev. 0.037 / 0.067 dB/m \\ \vspace{0.1cm}}\\
\hline 
\shortstack{\\ Praks et al. (2007) \cite{ar:praks2007}\\ \vspace{0.25cm}} &\shortstack{\\  Boreal \\ \vspace{0.25cm}}& \shortstack{\\ n/a \\0.1\\ \vspace{0.15cm}} & \shortstack{\\ Quad-pol  \\ VV\\ \vspace{0.15cm}} & \shortstack{\\ L  \\ X\\ \vspace{0.15cm}} & \shortstack{\\Not provided at L-band \\ InSAR approach \\ Assumed at X-band}\\
\hline 
Praks et al. (2007) \cite{pro:praks2007}& Boreal & \shortstack{\\0.14 \\0.09 \\0.13} & \shortstack{\\ HH \\ HV \\ VV } & \shortstack{\\ \quad \\X \\ \\\quad \\\quad} & \shortstack{\\ \vspace{0.15cm}\\ Estimated from \\HUTSCAT scatterometer \\$\theta_0$=0$^\circ$ }\\

 &  & \shortstack{\\ 0.9 (median)\\ \vspace{0.4cm}} & \shortstack{\\ VV\\ \vspace{0.4cm}} & \shortstack{\\ X\\ \vspace{0.4cm}} & \shortstack{\\ \\\quad \quad \\E-SAR, $\theta_0$ n/a \\ InSAR approach \\\emph{Small influence on} \\ \emph{height inversion}} \\


\hline
\end{tabular}
\end{center}
\caption{\small Overview of contributions employing, estimating and/or measuring 
the extinction coefficient (1/2)} \label{t:ext1}
\end{table}


\begin{table}[h!]
\renewcommand{\arraystretch}{1.0}
\scriptsize
\begin{center}
\begin{tabular}{|l|c|c|c|c|c|}
\Xhline{1.5pt}  \rowcolor{red_light}
{\bf Reference} & {\bf Forest type} & {\bf Extinction value (dB/m)} & {\bf Polarisation} &
{\bf Freq. Band} & {\bf Comments} \\ 
\Xhline{1.5pt}
\shortstack{\\ Garestier et al. (2008) \cite{ar:garestier08_x} \\ \vspace{0.3cm}}& \shortstack{\\Sparse \\pine forest\\ \vspace{0.3cm}} & \shortstack{\\1.6\\ \vspace{0.3cm}} & \shortstack{\\HH / HV\\ \vspace{0.3cm}} & \shortstack{\\X\\ \vspace{0.3cm}} & \shortstack{\\Tested range is 0.4 to 2 B/m \\for a HH/HV PolInSAR inversion \\ being 1.6 dB/m the mean value.\\ \emph{Limited influence of} \\ \emph{the mean extinction coefficient}}\\
%
\hline 
\shortstack{\\ Garestier et al. (2008) \cite{ar:garestier08_p} \\ \vspace{0.3cm}} & \shortstack{\\Maritime  \\ pine trees\\ \vspace{0.1cm}} & \shortstack{\\ 0.3-0.5\\ \vspace{0.3cm}} & \shortstack{\\ Quad-pol \\ \vspace{0.3cm}}& \shortstack{\\ P\\ \vspace{0.3cm}} &
\shortstack{\\Sensitivity analysis\\suggesting these \\values}\\
\hline 
\shortstack{\\ Hajnsek et al. (2009) \cite{ar:hajnsek_indrex_09}\\ \vspace{0.4cm}} & \shortstack{\\ Tropical \\ \vspace{0.4cm}}& \shortstack{\\ 0.3 \\ \vspace{0.4cm}}& \shortstack{\\ VV \\ \vspace{0.4cm}}& \shortstack{\\ X\\ \vspace{0.4cm}} & \shortstack{\\Estimation range is 0.1-0.9 dB/m \\ with a 0.3 dB/m mean value. \\ P- and L-band quad-pol \\ PolInSAR applied but \\ no extinction values provided.}\\
\hline
\shortstack{\\Neumann et al. (2010) \cite{ar:neumann2010} \\ \vspace{0.4cm}}& \shortstack{\\Temperate\\ \vspace{0.4cm}} & \shortstack{\\0-0.3 \\ \vspace{0.4cm}}&\shortstack{\\ Quad-pol \\ \vspace{0.4cm}}& \shortstack{\\L \\ \vspace{0.4cm}} & \shortstack{\\SB and MB PolInSAR inversion \\ Validation of extinction values \\ \emph{requires more information} \\ about forest structure.}\\

\hline Praks et al. (2012) \cite{ar:praks2012} & Boreal & \shortstack{\\0.18 \\0.21} & \shortstack{\\ HV \\ HH-VV } & \shortstack{\\ \quad \\L \\ \\\quad \\\quad} & \shortstack{\\ Estimates up to 0.7 dB/m \\InSAR approach }\\

&  & 0.4 & VV & X & \shortstack{\\\quad \\ Estimates up to 0.8 dB/m \\InSAR approach } \\ 
\hline 
\shortstack{\\Kugler et al. (2014) \cite{ar:kugler2014} \\ \vspace{0.3cm}}& \shortstack{\\Boreal, \\ temperate, \\and tropical\\ \vspace{0.05cm}} & \shortstack{\\n/a \\ \vspace{0.3cm}}& \shortstack{\\single-pol / dual-pol\\ \vspace{0.3cm}} & \shortstack{\\X \\ \vspace{0.3cm}}& \shortstack{\\Single- and dual-pol \\ PolInSAR inversion \\ Extinction is inverted \\ but not provided.}\\
\hline 
\shortstack{\\Kugler et al. (2015) \cite{ar:kugler2015} \\ \vspace{0.2cm}}& \shortstack{\\Boreal \\Temperate \\Tropical} & \shortstack{\\0-0.1 dB/m, \emph{rarely} $>$0.2 dB/m\\0-0.2 dB/, \emph{rarely} $>$0.5 dB/m\\0-0.2 dB/, \emph{rarely} $>$0.5 dB/m} & \shortstack{\\Quad-pol \\ \vspace{0.2cm}}& \shortstack{\\ L  \\ L  \\ P} & \shortstack{\\PolInSAR inversion\\ \vspace{0.2cm}}\\
\hline 
Chen et al. (2018) \cite{ar:chen2018} & \shortstack{\\ Dense mountainous \\temperate forests} & \shortstack{\\Mostly up\\to 0.3 dB/m} & HH & X & \shortstack{\\ InSAR inversion \\ Some $>$0.3 dB/m}\\
\hline
\shortstack{\\Tora\~no-Caicoya et al. (2016) \cite{ar:caicoya2016} \\ \quad \\ \quad \\ \quad \\ \quad \\ \quad} & \shortstack{\\Boreal\\ \quad \\ \quad \\ \quad\\ \quad \\ \quad} & \shortstack{\\0.1\\ \quad \\ \quad \\ \quad\\ \quad \\ \quad} & \shortstack{\\HH / VV\\ \quad \\ \quad \\ \quad\\ \quad \\ \quad}& \shortstack{\\X \\ \quad \\ \quad \\ \quad \\ \quad \\ \quad} & \shortstack{\\Tested range is 0-0.3 dB/m \\for a single-channel InSAR inversion \\ being 0.1 dB/m the best fit.\\ Dual-baseline approach \\also assumed 0.1 dB/m.\\\emph{Wide range can be used}}\\ 
\hline
Askne et al. (2017) \cite{ar:askne2017} & Boreal & 0.11-0.16 & VV & X & \shortstack{\\ Combined InSAR (IWCM) \\ and backscattering inversion}\\
\hline
\shortstack{\\Lei et al. (2018) \cite{ar:lei2018}\\ \quad\\ \quad \\ \quad \\ \quad \\ \quad \\ \quad \\ \quad \\ \quad} & \shortstack{\\Tropical (primary) \\ Tropical (secondary)\\ \quad \\ \quad \\ \quad \\ \quad \\ \quad \\ \quad \\ \quad} & \shortstack{\\ 0.05 \\ 0.1\\ \quad \\ \quad \\ \quad \\ \quad \\ \quad \\ \quad \\ \quad} & \shortstack{\\HH \\ \quad \\ \quad\\ \quad \\ \quad \\ \quad \\ \quad \\ \quad \\ \quad \\ \quad} & \shortstack{\\X \\ \quad \\ \quad\\ \quad \\ \quad \\ \quad \\ \quad \\ \quad \\ \quad \\ \quad} & \shortstack{\\ InSAR inversion \\ Large gaps \\ in the canopy \\ could explain \\low extinction \\ \cite{ar:askne,ar:treuhaft2009}}\\
%
\hline 
\shortstack{\\ \quad \\ \quad \\Cartus et al. (2019)  \cite{ar:cartus2019}\\ \quad\\ \quad \\ \quad \\ \quad \\ \quad \\ \quad \\ \quad \\ \quad} & \shortstack{\\Tropical \\ \quad\\ \quad \\ \quad \\ \quad \\ \quad \\ \quad \\ \quad \\ \quad}& \shortstack{\\0.125 / 0.5 \\ \quad\\ \quad \\ \quad \\ \quad \\ \quad \\ \quad \\ \quad \\ \quad} & \shortstack{\\HH, HV, VV \\ \quad\\ \quad \\ \quad \\ \quad \\ \quad \\ \quad \\ \quad \\ \quad}& \shortstack{\\ L  / C, X \\ \quad\\ \quad \\ \quad \\ \quad \\ \quad \\ \quad \\ \quad \\ \quad}& \shortstack{\\Assumed values but \\ not assured they are realistic \\ Same values for \\ all channels\\ \quad \\ \quad}\\
\hline 
\end{tabular}
\end{center}
\caption{\small (cont.) Overview of contributions employing, estimating and/or measuring 
the extinction coefficient (2/2).} \label{t:ext2}
\end{table}

\vspace{0.5cm}
After a careful review of these contributions (see Tables \ref{t:ext1} and \ref{t:ext2}) several comments are in order:

\begin{itemize}
\item Among all works reviewed only three contributions reported specific experimental measurements on the 
extinction coefficient (indicated by the grey coloured cells in Table \ref{t:ext1}). In all three cases, the L-band extinction coefficient presented similar values despite they all correspond to different forest types. Early studies presented in \cite{ar:kasischke89} and \cite{ar:way1990} reported values obtained through dedicated experiments as well but we were unable to find the corresponding research documents.

\item A noticeable number of works claim, either as an implicit or 
explicit conclusion, that the extinction value is not a critical parameter and it can adopt a variable value inside 
a certain interval. This seems completely justified in terms of the resulting accuracy of the final 
product (i.e. height or biomass) and more specifically when employing InSAR approaches where the absolute 
power is normalised to calculate the coherence. However, it is also reasonable to assume that a 0.1 dB/m 
difference (for instance from 0.2 to 0.3 dB/m) is bound to have a remarkable impact when using 
backscattering-based approaches for tall vegetation.

\item Significant penetration of the radar signal into the forest canopy has been suggested even at X-band due to large gaps in the vegetation cover. This effect seems to explain the apparent low extinction at this short wavelength. However, higher extinctions were also found in sparse forest (\cite{ar:garestier08_x}).

\item Nearly all contributions employing C-band data were focused on boreal forest where extinction ranges between 0.08 and 2 dB/m being in some cases similar to values either at L- or at X-band alike. On the other hand, only one contribution (\cite{ar:cartus2019}) treated the extinction issue in tropical areas at C-band.

\item For the cases focused on P-band studies, experimental measurements of extinction \cite{ar:thirion_COSMO} are noticeably  lower than the values derived from PolInSAR inversion \cite{ar:garestier08_p} for maritime pine trees. However, a third contribution also based on PolInSAR inversion in tropical forest \cite{ar:kugler2015} reported extinction values similar to those in \cite{ar:thirion_COSMO}.
\end{itemize}

Even though this diversity of results could be due to the different morphological and environmental conditions in every test site, it also draws the attention to the fact that a better knowledge on the extinction coefficient and its dependence on the sensor frequency would be required. Therefore, it is reasonable to think that further research specially focused on experimental measurements of the extinction coefficient is necessary.  

\section{Random Volume over Ground backscattering model }

The backscattering model employed in this report is directly derived from the physically-based framework employed by Treuhaft et al. \cite{ar:treuhaft,ar:treuhaft_siqueira} to derive the Random Volume over Ground (RVoG) model. It is noted that a similar approach was presented by Hensley et al. \cite{ar:hensley2014} but most of parameters were treated as fitting coefficients. However, our option is to keep the physical meaning of them in order to ascertain the role that (in particular) the extinction coefficient adopts. 

We focus on the case where the scattering mechanisms from the canopy, i.e. the volume component, and a disturbed (i.e. modified by direct scattering due to local topography and understory) double-bounce interaction between vegetation and soil control the overall radar response as a function of vegetation height. This assumption has been observed to approximate reasonably well the radar signature at P- \cite{ar:tebaldini2009} and L-band \cite{ar:yu_saatchi}. In addition, the resulting physical model can be transformed into a semi-empirical model accounting also for the direct surface backscatter by assuming independent mechanisms in a monostatic acquisition as it is also located at the ground level \cite{ar:bouvet2018}.

The RVoG backscattering model is represented as shown in Eq.(\ref{eq:model}) from the derivation in \cite{ar:treuhaft,ar:treuhaft_siqueira}:

\begin{equation}
P =  \frac{P_v}{2\sigma/\cos\theta_0} \cdot \left( 1 - e^{-\frac{2\sigma h_v}{\cos\theta_0}}\right) + P_{dbl} \cdot h_v \cdot e^{-\frac{2\sigma h_v}{\cos\theta_0}}
\label{eq:model}
\end{equation}

where $P_v$ and $P_{dbl}$ represent the backscattering power from volume and double-bounce mechanisms, respectively; $\sigma$ is the one-way extinction coefficient (Np/m); $h_v$ represents the volume height; and $\theta_0$ is the incidence angle. In order to diminish the complexity and treatment of the model (\ref{eq:model}) different parameters (i.e. volume and volume specular backscattering, ground roughness, volume and surface scatterers densities and so on) were grouped to define $P_v$ and $P_{dbl}$. The reader is referenced to \cite{ar:treuhaft_siqueira} for further details.

It is noted that model (\ref{eq:model}) can be alternatively expressed as in (\ref{eq:model2}), thus resembling the empirically-based models commonly employed for height (and biomass) retrieval:

\begin{equation}
P = a_1 \cdot \left( 1 - e^{-a_2 h_v}\right) + a_3 e^{-a_2 h_v}
\label{eq:model2}
\end{equation}

In (\ref{eq:model2}) some parameters have been grouped and renamed as follows:

\begin{align}
  a_1 & = P_v \frac{\cos\theta_0}{2\sigma} \nonumber \\
  a_2 & = \frac{2\sigma}{\cos\theta_0}  \label{eq:a_c} \\
	a_3 & = P_{dbl} \nonumber
\end{align}

Next, an assessment of model (\ref{eq:model}) is performed for some particular cases. As we will show, the main conclusion from this theoretical analysis is in agreement with the outcomes from \cite{ar:mermoz2015} regarding the decreasing trend of backscattering beyond the saturation point.

\subsection*{Model Assessment}

\begin{figure*}[h!]
 \centering
\begin{tabular}{c}
  \includegraphics[width=14cm]{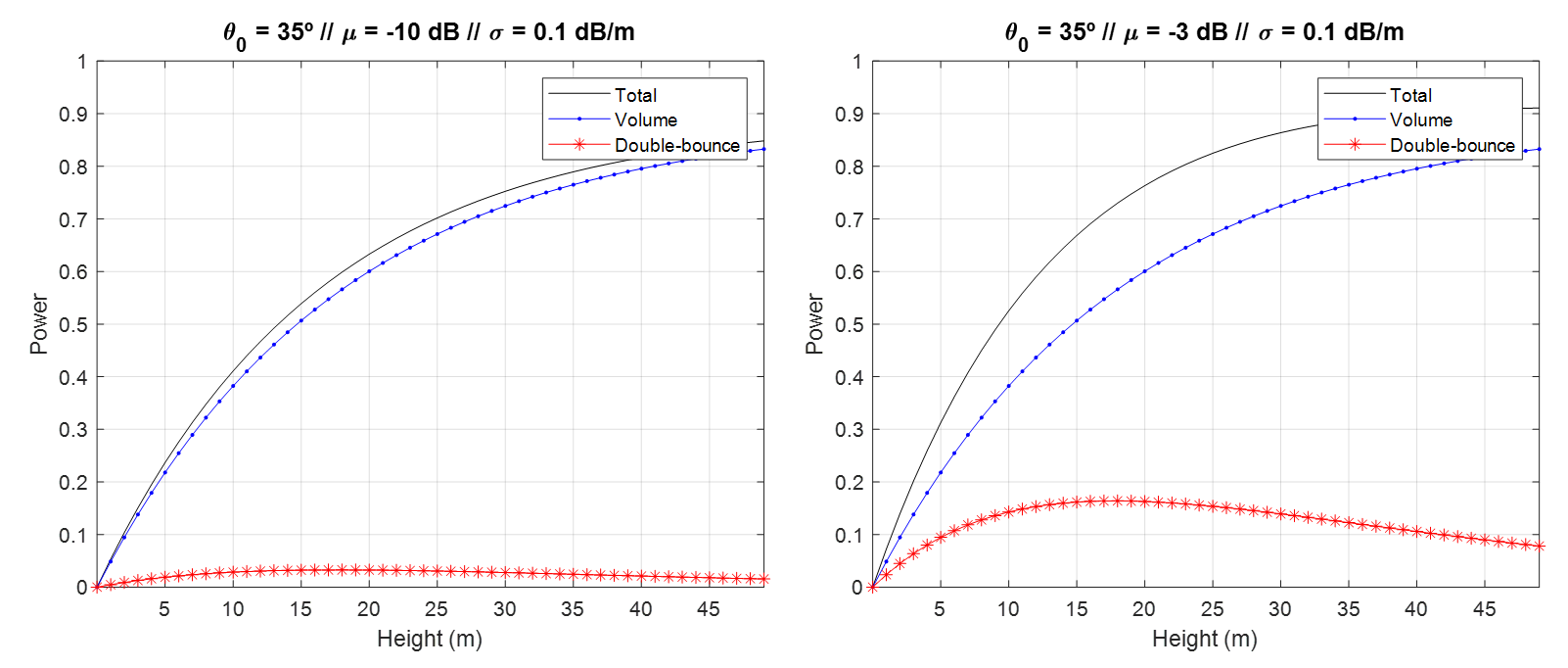}
\end{tabular}
\caption{RVoG backscattering model for an extinction coefficient of $\sigma$=0.1 dB/m and (left) $\mu$=-10 dB and (right) $\mu$=-3.}
\label{f:sim_0_1}
\end{figure*}

\begin{figure*}[h!]
 \centering
\begin{tabular}{c}
  \includegraphics[width=8cm]{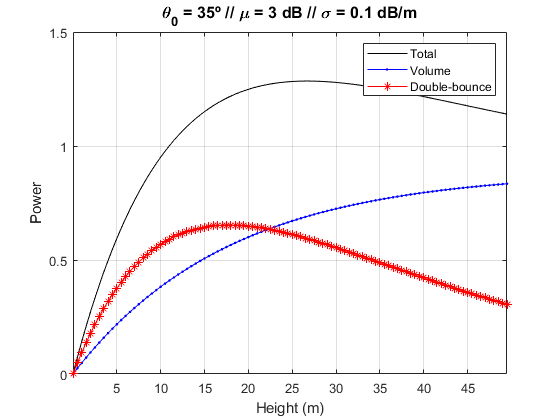}
\end{tabular}
\caption{RVoG backscattering model for an extinction coefficient of $\sigma$=0.1 dB/m and higher double-bounce than volume scattering, i.e. $\mu$=3 dB.}
\label{f:sim_0_1_mu_high}
\end{figure*}

Next, the backscattering model (\ref{eq:model}) is plotted as a function of volume height $h_v$ for particular values of extinction $\sigma$ and volume and double-bounce powers. Figure \ref{f:sim_0_1} displays two cases for a low and a non-negligible ground-to-volume ratio, respectively, with the same extinction $\sigma = 0.1$ dB/m and incidence angle of 35$^\circ$. Parameter $\mu$ at the top of both figures represents this ratio defined as $\mu = P_{dbl}/P_v$. Total backscattering (i.e. model in (\ref{eq:model})) and each isolated scattering contribution are represented. In this case the volume contribution (the asymptotic component) dominates the backscattering trend over the non-asymptotic double-bounce component, whose functional form is $k_1 x e^{-k_2 x}$. This plot would suggest that the double-bounce contribution could contribute to shape the variation of the total backscattering level as a increasing-decreasing behaviour with a clear maximum at certain height $h_v$. Actually, if we now assume a $\mu$=3 dB, i.e. the double-bounce power doubles the volume power, then Figure \ref{f:sim_0_1_mu_high} shows that the non-asymptotic component dominates now the total response and a maximum backscattering level can be identified. It is noted, however, that this situation is not a realistic one as the distinctive feature in forests is that the volume backscattering is higher than the ground components.

\begin{figure*}[h]
 \centering
\begin{tabular}{c}
  \includegraphics[width=14cm]{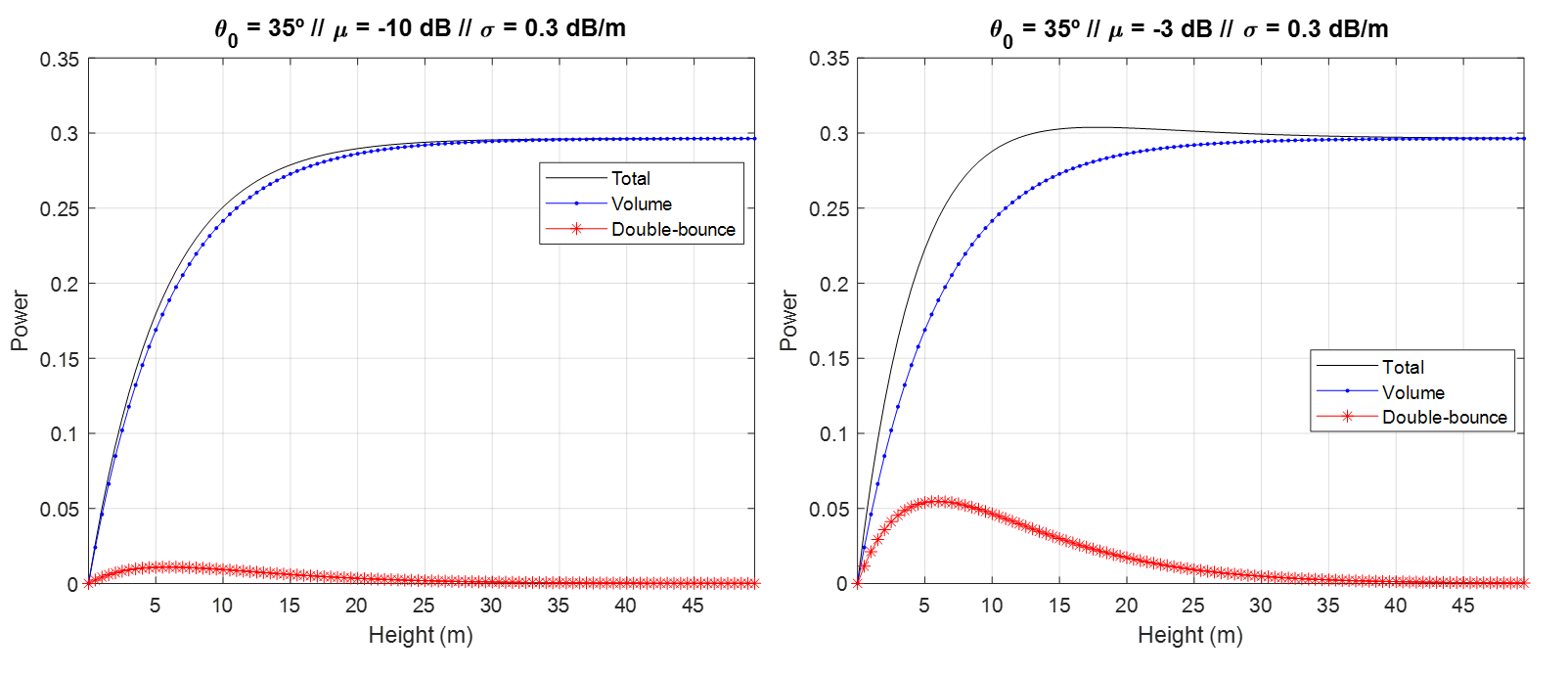}
\end{tabular}
\caption{RVoG backscattering model for an extinction coefficient of $\sigma$=0.3 dB/m and (left) $\mu$=-10 dB and (right) $\mu$=-3.}
\label{f:sim_0_3}
\end{figure*}

Besides the $\mu$ parameter the variation rate of the model also depends on the extinction coefficient $\sigma$. Keeping the $\mu$ values below 0 dB (i.e. volume component higher than the double-bounce) and increasing the extinction to 0.3 dB/m leads to the plot shown in Figure \ref{f:sim_0_3}. Increasing the extinction does not change the asymptotic behaviour in case of a very low double-bounce component ($\mu$=-10 dB). However, in case the double-bounce could not be considered as a negligible contribution ($\mu$=-3 dB) then this higher extinction leads to an increasing-decreasing trend with a maximum backscattering value at certain $h_v$ value.

According to the present model and the increasing-decreasing behaviour it exhibits for certain cases, i.e. when the asymptotic term is not the dominant effect, it is possible to calculate the $h_v$ value at which the backscattering is maximum:

\begin{equation}
\frac{\partial P}{\partial h_v} = 0 \label{eq:deriv}
\end{equation}

where $P$ is the RVoG backscattering model in (\ref{eq:model}).

Solving (\ref{eq:deriv}) for $h_v$ leads to expression (\ref{eq:hv}) where the volume height $h^{sat}_v$ corresponding to the maximum backscattering  is as follows:

\begin{equation}
h^{sat}_v = \frac{\cos\theta_0 \cdot (P_v + P_{dbl})}{2 \sigma P_{dbl}} \label{eq:hv}
\end{equation}

Considering that the ground-to-volume ratio is defined as $\mu = P_{dbl}/P_v$, then (\ref{eq:hv}) can be expressed as (\ref{eq:hv2}):

\begin{equation}
h^{sat}_v = \frac{\cos\theta_0 \cdot (1 + \mu)}{2 \sigma \mu} \label{eq:hv2}
\end{equation}

The relationship shown in Eq. (\ref{eq:hv2}) expresses in a compact way how extinction, volume height at the maximum backscattering and the ground-to-volume ratio are interrelated according to this type of backscattering models and assuming that it is plausible that backscattering decreases after reaching a maximum value for certain volume height \cite{ar:mermoz2015}. 

From (\ref{eq:hv2}) maps relating all three parameters can be obtained. Expressions for $\sigma = f(h^{sat}_v,\mu)$ or alternatively $\mu = f(h^{sat}_v,\sigma)$ are obtained from (\ref{eq:hv2}) as follows: 

\begin{align}
  \sigma & = \frac{\cos\theta_0 \cdot (1 + \mu)}{2 h^{sat}_v \mu} \label{eq:s_f}\\
	 & \nonumber\\
  \mu & = \frac{\cos\theta_0}{2\sigma h^{sat}_v - \cos\theta_0}
	\label{eq:mu_f}
\end{align}

\begin{figure*}[h!]
 \centering
\begin{tabular}{c}
  \includegraphics[width=8cm]{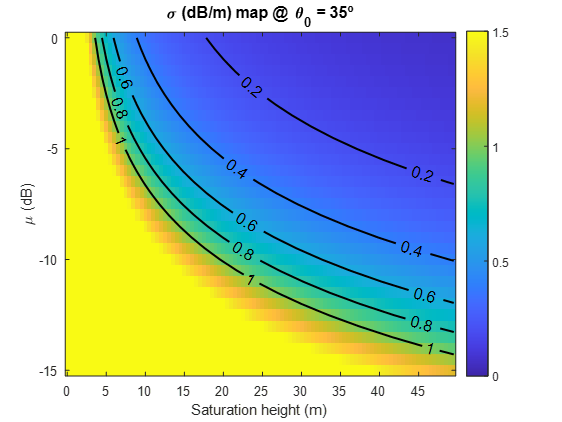}
\end{tabular}
\caption{RVoG backscattering model: Values of extinction coefficient (dB/m) as a function of volume height at the maximum backscattering and the ground-to-volume ratio for a 35$^\circ$ incidence angle.}
\label{f:s_map}
\end{figure*}

\begin{figure*}[h!]
 \centering
\begin{tabular}{c}
  \includegraphics[width=8cm]{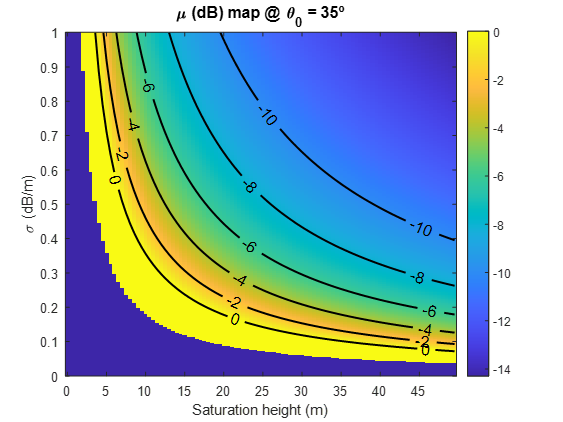}
\end{tabular}
\caption{RVoG backscattering model: Values of $\mu$ (dB) as a function of volume height at the maximum backscattering and the extinction coefficient for a 35$^\circ$ incidence angle.}
\label{f:mu_map}
\end{figure*}

Maps corresponding to expressions (\ref{eq:s_f}) and (\ref{eq:mu_f}) for a 35$^\circ$ incidence angle are displayed in Figures \ref{f:s_map} and \ref{f:mu_map} where contour plots overlap the corresponding maps.

In case of expression (\ref{eq:s_f}), yellow area in Figure \ref{f:s_map} represent unrealistic values of $\sigma$ higher than 1 dB/m for forest not reported in the frequency bands commonly employed for radar remote sensing of vegetation. However, the range of feasible $\sigma$ values enlarges (blue to greenish colors, i.e. null extinction to around 0.8 dB/m) as the height at maximum backscattering (i.e. $h^{sat}_v$) increases thus allowing a wider range of $\mu$ values. Therefore, this figure also explains how the model in (\ref{eq:model}) constraints the possible extinction values according to certain intervals of $\mu$ and $h^{sat}_v$. 

Alternatively, Figure \ref{f:mu_map} represents the map derived from (\ref{eq:mu_f}) where the range of $\mu$ values is depicted. From this plot, a canopy represented by a 0.3 dB/m extinction should exhibit a maximum backscattering at a 29.5 m in order to assume (or retrieve) a $\mu$ approximately equal to -6 dB. Note this condition becomes even stricter for a 0.1 dB/m extinction. Hence, a constraint on the minimum ground double-bounce return measured by the radar must be observed in order to employ the two-component model in (\ref{eq:model}) for parameter inversion. 
\newpage
It is also noted in Figure \ref{f:mu_map} that the dark blue homogeneous area located close to $x$ and $y$ axes corresponds to $(\sigma,h_v)$ pairs producing negative non-feasible values of $\mu$. By considering the denominator of expression (\ref{eq:mu_f}) it is seen that the following condition must be fulfilled in order to avoid negative values of $\mu$:

\begin{equation}
  \frac{\cos\theta_0}{2\sigma \cdot h^{sat}_v - \cos\theta_0} > 0 \quad \longrightarrow \quad \sigma \cdot h^{sat}_v > \frac{\cos\theta_0}{2}
	\label{eq:cond}
\end{equation}

Condition shown in (\ref{eq:cond}) can be also directly obtained from the term between parentheses in (\ref{eq:model}) since it arises as a consequence of forcing the volume component not to take negative values. Therefore, the theoretical backscattering signature as a function of height can be modeled as an increasing-decreasing trend under this constraint. On the other hand, when (\ref{eq:cond}) is not fulfilled then it leads to an asymptotic behaviour and only the first term in (\ref{eq:model}) related to the volume scattering must be considered.

\subsection*{Comments on parameter inversion}

By making use of an a priori knowledge of the forest height and a regression approach, the model parameters, $\sigma$, $P_v$, and $P_{dbl}$ can be estimated. Retrieving positive values of these unknowns (together with the common statistical indicators to assess the quality of the fitting) would confirm the validity of the non-asymptotic model (\ref{eq:model}). Otherwise, if no upward-downward trend is exhibited by the data then the regression procedure will yield very low or even negative values for the ground power $P_{dbl}$. Then, a more simplified model only accounting for the asymptotic component (i.e. the volume contribution) as shown in Eq.(\ref{eq:model_a}) will be employed.

\begin{equation}
P =  \frac{P_v}{2\sigma/\cos\theta_0} \cdot \left( 1 - e^{-\frac{2\sigma h_v}{\cos\theta_0}}\right)
 \label{eq:model_a}
\end{equation}

The aforementioned observations are illustrated below by applying the regression analysis by using the equivalent model (\ref{eq:model2}) where unkowns $a_1$, $a_2$, and $a_3$ were defined as shown in (\ref{eq:a_c}). To this aim we made use of an ALOS-1 PALSAR image acquired on April 21st 2009 (product ALPSRP172570900-P1.1) over Howland Forest, Maine, Northeastern United States. The SAR image was downloaded from the Alaska SAR Facility Vertex data portal from NASA's EOSDIS. 

Howland forest is a boreal transitional forest mostly dominated by mixed spruce, hemlock, aspen, and birch stands. The Lidar data were collected by LVIS sensor \cite{ar:lvis} at a nominal 20 m grid (product LVIS\_US\_ME\_2009\_VECT\_20100328). The RH100 metric was employed in this study. More details on the region are given in \cite{ar:lei2019}.

Regression results were computed by means of a numerical optimisation carried out in terms of a non-linear least squares problem solved by using the \emph{lsqcurvefit} tool from Matlab. Despite this solver allows to set upper and lower bounds for limiting the interval of solutions this option was not employed for these results. The values for initial guess of all three unkowns $a_1$, $a_2$, and $a_3$ for initializing the procedure are set according to the following considerations:

\begin{enumerate}
\item $a_1$: As it contains the effect of both $P_v$ and $\sigma$, its value is set to half the average backscattering power of all plots considered for the regression.
\item $a_2$: One hundred values are randomly taken from a 0.05-0.4 dB/m extinction interval. Then, the corresponding $a_2$ values (see Eq.(\ref{eq:a_c})) are calculated according also to the incidence angle (24$^\circ$). Thus the regression procedure is repeated one hundred times. The final extinction value is provided in terms of the average of all solutions and the corresponding standard deviation.
\item $a_3$: Assuming that the ground component is lower than the volume contribution, its value is set to five times lower than $a_1$.
\end{enumerate}

\begin{figure*}[h!]
 \centering
\begin{tabular}{cc}
  \includegraphics[width=8cm]{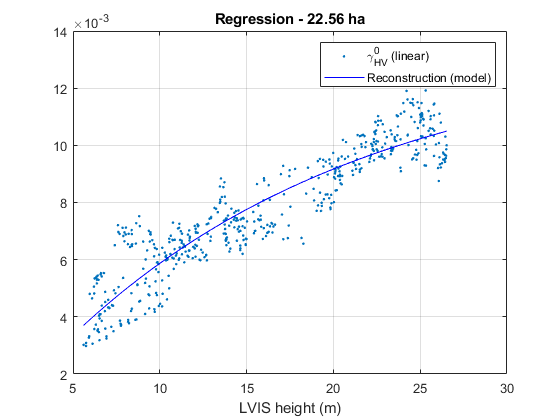}  &   \includegraphics[width=8cm]{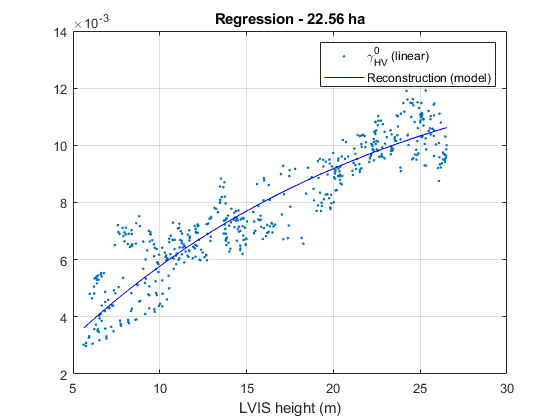}\\
	a) & b)
\end{tabular}
\caption{Regression results from ALOS-1 data in Howland forest employing a) The non-asymptotic model (\ref{eq:model2}), equivalent to (\ref{eq:model}), and b) The asymptotic model (\ref{eq:model_a}). Results for $a_1$ and $a_2$ for both cases are nearly the same (see Tables \ref{t:reg_nasym} and \ref{t:reg_asym})}
\label{f:reg_nasym}
\end{figure*}

Figure \ref{f:reg_nasym}.a displays the regression results considering the non-asymptotic model (\ref{eq:model}) equivalent to (\ref{eq:model2}). Table \ref{t:reg_nasym} presents the results for $P_v$, $\sigma$, and $P_{dbl}$ derived from $a_1$, $a_2$, and $a_3$. The squared 2-norm of the residual at the final solution ranges between 5.33$\cdot10^{-4}$ and 5.47$\cdot10^{-4}$ for the whole set of realizations which shows a good performance from the mathematical viewpoint. However, the retrieved value for $a_3$ which represents the double-bounce backscattering $P_{dbl}$ is negative. Therefore, this results would suggest that the present data do not follow the upward-downward signature as a function of height but an asymptotic behaviour instead. 

\begin{table}[h]
\begin{center}
\begin{tabular}{|c|c|c|c|c|}
\hline
$P_v$    & \begin{tabular}[c]{@{}l@{}}$\sigma$ (dB/m)\\ Avg. / Std. dev.\end{tabular} & $P_{dbl}$      & Min. Resnorm   & Max. Resnorm   \\
\hline 
0.00078 & 0.1168 / 7.3 $10^{-4}$   & -1.34 $10^{-5}$ & 5.33 $10^{-4}$ & 5.47 $10^{-4}$ \\
\hline
\end{tabular}
\end{center}
\caption{\small Parameters estimated by using the non-asymptotic model (\ref{eq:model}) expressed as in (\ref{eq:model2}).}
\label{t:reg_nasym}
\end{table}

According to the results shown in Table \ref{t:reg_nasym} the regression is again carried out but assuming the simplified asymptotic model in (\ref{eq:model_a}) where only $a_1$ and $a_2$ must be retrieved. Results are shown in Figure \ref{f:reg_nasym}.b and Table \ref{t:reg_asym}. As shown, the performance is the same as in the non-asymptotic case but ignoring the double-bounce term. In particular, the extinction value estimate shown in Table \ref{t:reg_asym} is in partial agreement with estimates reported in some previous works focused on boreal forest, not only at L-band but also at C- and X-band as presented in Tables \ref{t:ext1} and \ref{t:ext2}.

\vspace{0.2cm}
As a final remark it is noted that environments where the increasing-decreasing trend is confirmed by measurements would allow the retrieval of the vegetation height $h^{sat}_v$ at the maximum backscatter from estimates of $\sigma$, $P_v$ and $P_{dbl}$ previously obtained by applying the regression analysis on Eq.(\ref{eq:model}). This case would allow to further investigate the possibility of enlarging the height estimation range on the basis of the different variation rates of backscattering below and above the $h^{sat}_v$ limit.

%
%
%
%

\begin{table}[h]
\begin{center}
\begin{tabular}{|c|c|c|c|}
\hline
$P_v$    & \begin{tabular}[c]{@{}l@{}}$\sigma$ (dB/m)\\ Avg. / Std. dev.\end{tabular} & Min. Resnorm   & Max. Resnorm   \\
\hline 
0.00074 & 0.1045 / 6.53 $10^{-4}$   & 5.32 $10^{-4}$ & 5.32 $10^{-4}$ \\
\hline
\end{tabular}
\end{center}
\caption{\small Parameters estimated by using the asymptotic model (\ref{eq:model_a}).}
\label{t:reg_asym}
\end{table}

\section{Summary}
\label{s:disc}

In this report two issues have been addressed concerning the role of extinction coefficient in radar backscattering-based models. Firstly, a review of a number of works has revealed a diversity of extinction values reported in the literature being unclear whether a direct and measurable relationship exists among the extinction values, the frequency band and the forest type. A number of works have realistically claimed that the extinction coefficient is not a critical parameter for a practical implementation of height and biomass inversion approaches. However, it may seem that this variability could possibly be linked to a certain lack of understanding of this parameter and its dependence on forest structural parameters as suggested also in several recent contributions \cite{ar:mermoz2015,ar:joshi2017,ar:cartus2019}. Therefore, further experimental studies specifically focused on radar signals interaction with vegetation (\cite{ar:albinet2012,pro:monteith2018}) are required.

On the other side, the formulation of the RVoG model \cite{ar:treuhaft,ar:treuhaft_siqueira} has been considered as an alternative option for backscattering analysis from vegetation. A RVoG version for backscattering modeling is obtained by ignoring in its original derivation the spatial diversity due to interferometric observations and assuming two dominant contributions, i.e. volume and double-bounce components. Since this is a model whose formulation only relies on physical foundations through the distorted Born approximation, then the physical meaning of all 
different parameters involved in the original formulation is maintained while keeping the model simple enough. Obviously, we do not claim that a physically-based model must be unconditionally selected instead of an empirical or semi-empirical one. These alternative approaches have been the logical and successful work-around to circumvent the complexity of the scattering processes involved in radar interaction with vegetation. Nevertheless, the use of physically-based models can provide complementary information in the analysis of inconsistencies between the expected behaviour of a physical process and the experimental measurements. In the present report, conclusions drawn from the analysis of the theoretical RVoG backscattering model points to an increasing-decreasing trend exhibited by backscattering as a function of height, thus existing some sensitivity beyond the saturation point. Backscattering reaches a maximum value for certain vegetation height which depends on the extinction coefficient and the ground-to-volume ratio. This observation is in agreement with statements made in \cite{ar:mermoz2015} where both electromagnetic simulations and experimental analysis were employed to bring the attention on the increasing-decreasing trend of L-band HV backscattering as a function of biomass for a tropical forest scenario. It is noted, however, that the sensitivity beyond the maximum backscattering is weak, then reasonable doubts arise on its practical implications \cite{ar:lucas2010,ar:yu_saatchi,ar:rodriguez2019} mainly because of the impact of the different sources of uncertainties on the data. Nevertheless, we believe the present discussion provides sufficient elements which suggest that further investigation on the relationship between backscattering and forest height can be done. An in-depth analysis on the consistency of these observations is currently underway supported by the existing data sets available. 

\vspace{2cm}

\section*{Acknowledgements}
Lidar data sets were provided by the Land, Vegetation and Ice Sensor (LVIS) team in Code 61A at NASA Goddard Space Flight Center with support from the University of Maryland, College Park. 

\vspace{2cm}

\bibliographystyle{unsrt}
\bibliography{mycites2,otherabrv,IEEEabrv}

\end{document}